 \documentclass[final,5p,times,twocolumn]{elsarticle}
\usepackage{dblfloatfix}    % To enable figures at the bottom of page
\usepackage{hyperref}
\usepackage{hypcap}
\usepackage{amssymb}
\usepackage{upgreek}
\newcommand{\+}{\hspace{.1em}}
\newcommand{\eV}{e\hspace{-.1em}V }
\newcommand{\eVd}{e\hspace{-.07em}V}

\journal{Nuclear Physics B}

\begin{document}

\begin{frontmatter}

%% Title, authors and addresses

%% use the tnoteref command within \title for footnotes;
%% use the tnotetext command for theassociated footnote;
%% use the fnref command within \author or \address for footnotes;
%% use the fntext command for theassociated footnote;
%% use the corref command within \author for corresponding author footnotes;
%% use the cortext command for theassociated footnote;
%% use the ead command for the email address,
%% and the form \ead[url] for the home page:
%% \title{Title\tnoteref{label1}}
%% \tnotetext[label1]{}
%% \author{Name\corref{cor1}\fnref{label2}}
%% \ead{email address}
%% \ead[url]{home page}
%% \fntext[label2]{}
%% \cortext[cor1]{}
%% \address{Address\fnref{label3}}
%% \fntext[label3]{}

\title{Second generation of portable gamma camera based on Caliste CdTe hybrid technology}

%% use optional labels to link authors explicitly to addresses:
%% \author[label1,label2]{}
%% \address[label1]{}
%% \address[label2]{}

\author[1]{MAIER Daniel}
\author[1]{BLONDEL Claire}
\author[1]{DELISLE Cyrille}
\author[1]{LIMOUSIN Olivier}
\author[1]{MARTIGNAC J\'{e}r\^{o}me}
\author[1]{MEURIS Aline}
\author[1]{VISTICOT Fran\c{c}ois}
\author[1,6]{DANIEL Geoffrey}
\author[2]{BAUSSON Pierre-Anne}
\author[2]{GEVIN Olivier}
\author[3]{AMOYAL Guillaume}
\author[3]{CARREL Fr\'{e}d\'{e}rick}
\author[3]{SCHOEPFF Vincent}
\author[4]{MAH\'{E} Charly}
\author[5]{SOUFFLET Fabrice}
\author[5]{VASSAL Marie-C\'{e}cile}

\address[1]{CEA Saclay, DRF/IRFU/D\'{e}partement d'Astrophysique, 91191 Gif-sur-Yvette Cedex, France}
\address[2]{CEA Saclay, DRF/IRFU/D\'{e}partement d'\'{E}lectronique, des D\'{e}tecteurs et d'Informatique pour la Physique, 91191 Gif-sur-Yvette Cedex, France}
\address[3]{CEA Saclay, LIST/Sensors and Electronic Architectures Laboratory, 91191 Gif-sur-Yvette Cedex, France}
\address[4]{CEA, Direction de l'\'{E}nergie Nucl\'{e}aire, 30207 Bagnols-sur-C\`{e}ze, Marcoule, France}
\address[5]{3D plus, 408 rue H\'{e}l\`{e}ne Boucher, 78530 Buc, France}
\address[6]{\'{E}cole Nationale des Ponts et Chauss\'{e}es, 6-8 avenue Blaise Pascal, Cit\'{e} Descartes, 77455 Marne-la-Vall\'{e}e, France}

\begin{abstract}
In the framework of a national funded program for nuclear safety, a first prototype of portable gamma camera was built and tested. It integrates a Caliste-HD CdTe-hybrid detector designed for space X-ray astronomy coupled with a new system-on-chip based acquisition system (FPGA and ARM microprocessor) and thermo-electrical coolers for a use at room temperature. The complete gamma part of the camera fits in a volume of $\medmuskip=0.5mu 15\+ \times \+15\+ \times \+40\,\mathrm{cm}^3$ for a mass lower than 1\,kg and a power consumption lower than 10\,W. Localization and spectro-identification of radionuclides in a contaminated scene were demonstrated during several test campaigns. A new generation of system is under development taking into account feedback experience from in-situ measurements and integrating a new generation of sensor cost-optimized by industrial applications called Caliste-O. Caliste-O holds a $\medmuskip=0.5mu 16\+ \times \+16$ pixel detector of \mbox{$\medmuskip=0.5mu 14\+ \times \+14\,\mathrm{mm}^2$} and 2\,mm thick with 8 full-custom front-end IDeF-X HD ASICs. Two prototypes were fabricated and tested. The paper will present the results of in-situ measurements with the first gamma camera, the spectroscopic performance of Caliste-O and the design of the second generation of gamma camera which aims for real time imaging and spectro-identification.

\end{abstract}

\begin{keyword}
ORIGAMIX \sep CALISTE-HD \sep CALISTE-O \sep Compton camera \sep gamma camera \sep CdTe
%% keywords here, in the form: keyword \sep keyword

%% PACS codes here, in the form: \PACS code \sep code

%% MSC codes here, in the form: \MSC code \sep code
%% or \MSC[2008] code \sep code (2000 is the default)

\end{keyword}

\end{frontmatter}

%\linenumbers
%\modulolinenumbers[5]

%% main text
\section{Introduction}
\label{sec:intro}
Since Chernobyl and more recently the Fukushima nuclear accident, nuclear safety became a major issue. Localization and spectro-identification of radioactive sources can give crucial information about the current accident stage and how to act on it. Besides, astrophysics does ask for spectro-imaging instruments development for years, and now such technologies are mature and ready to serve other domains. The ORIGAMIX (lOcalisation de la Radioactivit\'{e} par Imageurs GAMma pIXellis\'{e}s) project is funded by a French nuclear safety project call from the \textit{Investissements d'Avenir} national program. The first phase of the project consists in developing a portable gamma camera based on technological solutions coming from detector R\&D for astrophysics and testing it in real conditions. The second phase consists in designing a new nuclear safety camera with commercial grade solutions in the prospect of a large industrial application. 

\section{First prototype of portable gamma camera based on \mbox{Caliste-HD}}
\label{sec:cam1}
Caliste-HD \citep{CalisteHD, Dubos_2013} is a hybrid CdTe detector module developed for large focal plane arrays in space telescopes for high energy astronomy. It integrates a 256-pixel Schottky CdTe detector coupled with 8 full-custom front-end ASICs called IDeF-X HD. The attractive features of this device for nuclear safety applications are: its high spectral performance ($<\!\!1$\,keV FWHM at 60\,keV at $-10^\circ$C), its imaging capabilities ($\medmuskip=1mu 16\+ \times \+16$ pixels of $580\,\upmu$m pitch), its compact design ($\medmuskip=1mu 10\+ \times \+10\+ \times \+18\,\mathrm{mm}^3$ with the front-end electronics), and its low power (0.2\,W in total).

\begin{figure}[htb]
   \capstart
   \centering
   \includegraphics[width = 0.833\linewidth]{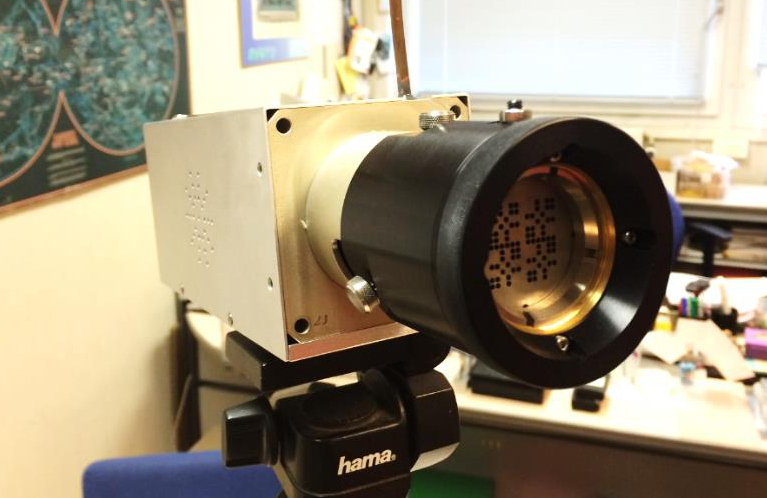}
   \caption{First prototype of ORIGAMIX gamma camera based on Caliste-HD detector and equipped with a MURA coded-mask (rank 7, $50^\circ \times \+50^\circ$ fully coded field of view). The full equipment (without the optical camera) stands in \mbox{$\medmuskip=1mu 15\+ \times \+15\+ \times \+40\,\mathrm{cm}^3$} for a mass lower than 1\,kg.}
   \label{fig:cam}
\end{figure}

\begin{figure*}[tb]
   \capstart
   \centering
   \includegraphics[width = \linewidth]{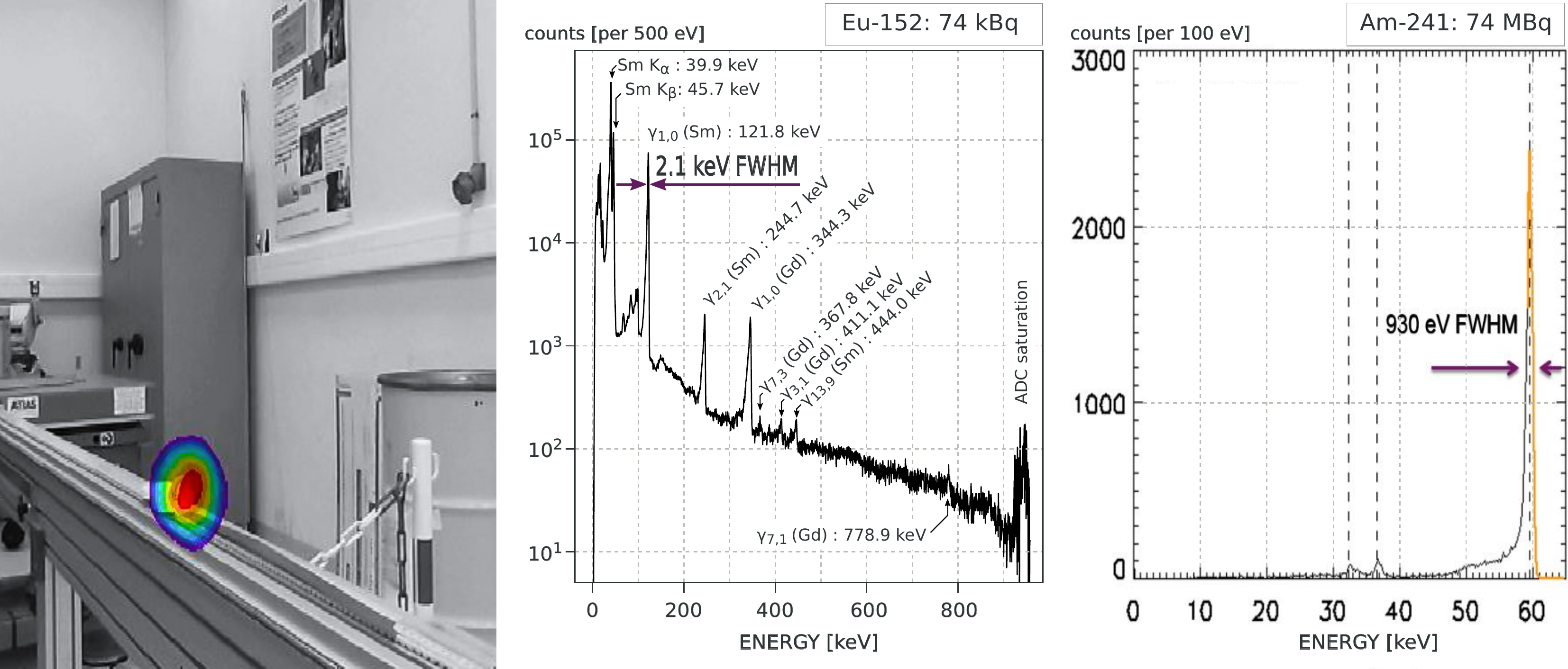}
   \caption{Experimental results obtained with the first prototype of ORIGAMIX gamma camera. \textbf{Left:} 
Superimposition of a gamma image ($^{152}$Eu source) obtained with a MURA coded mask (4\,mm W, rank 7, $50^\circ \times 50^\circ$ fully coded field of view) and an optical image. For the image reconstruction, no restriction on the photon energy was applied. \textbf{Center:} $^{152}$Eu spectrum obtained within an exposure of 4\,h in 5\,mm distance; lowest gain mode. In the presented configuration the energy limit was set by the analog-to-digital converter (ADC) and not by the ASIC. \textbf{Right:} Identification of an $^{241}\!\mathrm{Am}$ source (spectrum obtained within an exposure of 150\,s in 1\,m distance; highest gain mode). 
The $^{241}\!\mathrm{Am}$ spectrum is not representative of the spectral performance of the CALISTE detector at low energies due to the encapsulation (stainless steel package) of the source.
}
   \label{fig:campagne}
\end{figure*}

A mechanical structure with a sealed housing and removable coded-aperture mask, an acquisition system based on a system-on-chip (readout FPGA and ARM processor), and a cooling system with thermoelectric coolers were designed to integrate a Caliste-HD in a portable gamma camera (See Fig.\,\ref{fig:cam}). 
 
This system was tested in the lab and in-situ conditions. A $^{152}$Eu source of 10\,MBq activity placed at one meter distance from the camera ($1.2\,\upmu$Sv/h) was localized with $7^\circ$ angular resolution with the 4\,mm thick tungsten MURA mask (see Fig.\,\ref{fig:campagne}, left). An $^{241}\!\mathrm{Am}$ source of 74\,MBq activity at one meter distance is unambiguously identified in 400\,ms with only 30\,events thanks to the good energy resolution (below 1\,k\eV FWHM at 60\,k\eVd). In the same run the first 100~photons (1.2\,s exposure) were necessary to locate the point source in the decoded image. Figure~\ref{fig:campagne} (right) shows the spectrum after 150\,s. The source was encapsulated in a stainless steel package which absorbs the low energy lines.

\section{Second prototype of portable gamma camera based on \mbox{Caliste-O}}
\label{sec:cam2}
\begin{figure}[htb]
   \capstart
   \centering
   \includegraphics[width = 0.59\linewidth]{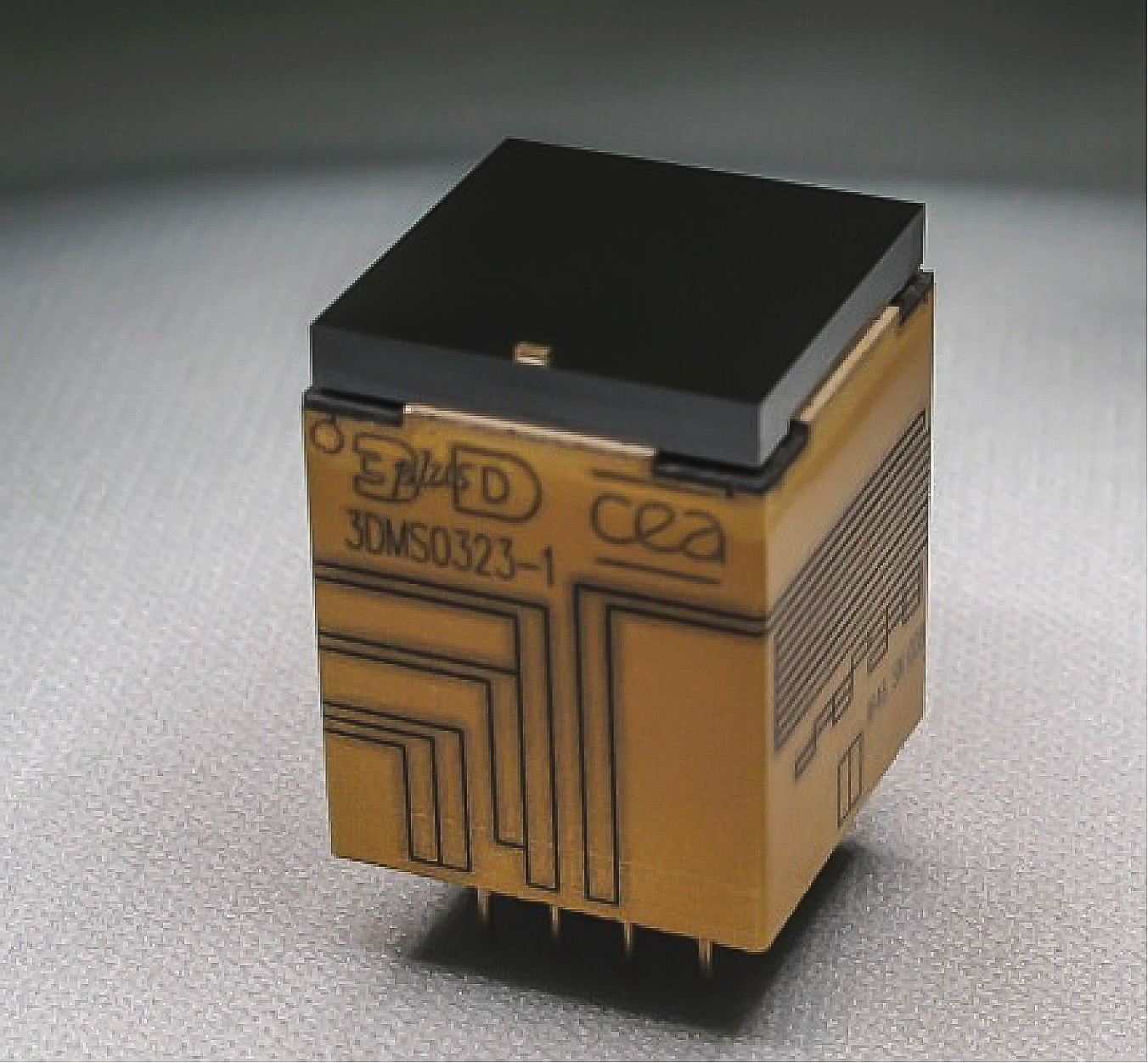}
   \caption{Caliste-O detector module: each of the 256 pixels of the CdTe detector is connected to a spectroscopic channel of full-custom ASICs. The resulting 3D hybrid component has a volume of $\medmuskip=1mu 15\+ \times \+15\+ \times \+18.7\,\mathrm{mm}^3$ for a total power of 0.2\,W.}
   \label{fig:module-o}
\end{figure}

A new generation of CdTe hybrid detector was designed and produced to better match safety nuclear applications: a CdTe pixel sensor with an increased detection area and crystal thickness in order to improve the detection efficiency. The crystal volume is $\medmuskip=1mu 14.2\+ \times \+14.2\+ \times \+2.0\,\mathrm{mm}^3$. Furthermore, simplified processes for fabrication and testing are applied for this commercial grade product. The result is called Caliste-O and is illustrated in Fig.\,\ref{fig:module-o}. The characterization of Caliste-O is presented in the following. A new mechanical structure has been designed to fit with the new dimensions of the module ($\medmuskip=1mu 15\+ \times \+15\+ \times \+18.7\,\mathrm{mm}^3$).

In parallel, some developments are on-going to improve the acquisition system based on experience feedback and to reduce the cost of the full system. A commercial and compact high voltage system will be procured. Electrical and tightness interfaces are made more reliable. Enhanced on-board data processing and user-friendly graphical interfaces are under study.

A first characterization of Caliste-O was performed in laboratory conditions (vacuum vessel, external cooling system, dedicated readout electronics) using the sources $^{241}\!\mathrm{Am}$, $^{137}$Cs, and~$^{60}$Co.

\subsection{$^\mathit{241}\!Am$ measurement}
\begin{figure*}[htb]
   \capstart
   \centering
   \includegraphics[width = 1\linewidth]{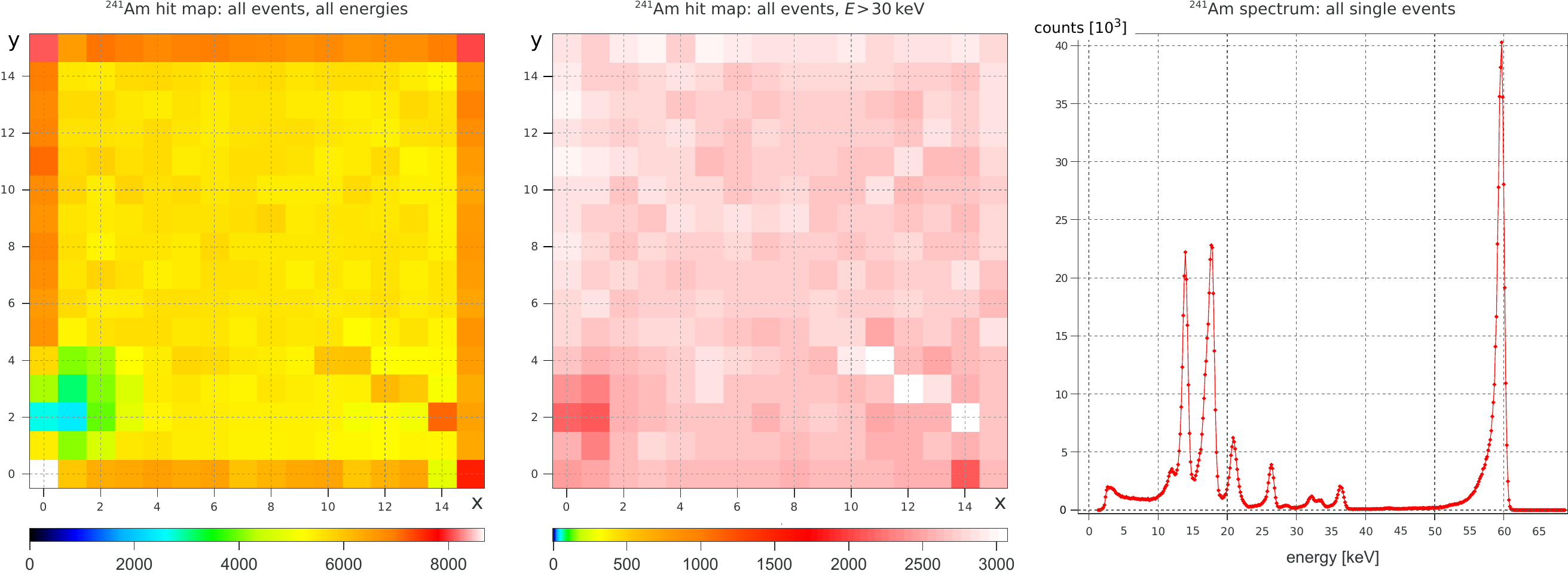}
   \caption{Total hit map (left), hit map with event selection $E >30$\,k\eV (center), and spectrum (right) of a measurement with $^{241}\!\mathrm{Am}$. The energy resolution was measured with $\Delta E = 927\,$eV FWHM at 60\,keV. Operating conditions: $U=500\,$V, $T=-20^\circ$C, highest gain mode.}
   \label{fig:am241}
\end{figure*}

Figure \ref{fig:am241} shows the result of a $^{241}\!\mathrm{Am}$ measurement. All 256 pixel are working as expected and most of the pixels show a uniform total flux response within $\pm 7.5\,\%$ maximum. Only a few pixels deviate from this average response and need further explanation.
\begin{itemize}
   \item The reduced counting rate around the position $x=1$, $y=2$ is caused by additional X-ray absorption of the depletion voltage contact, which was realized with a metallic spring pressing on a silicon pad on top of the crystal cathode. This feature is strongly reduced at higher energies. In the final integration of Caliste-O, this contact will be realized with a bonding to a dedicated pad on the top edge of the crystal which will cause no absorption as this position is above the guard ring and not above the pixels.
   \item The enhanced counting rate on the edges of the module are not caused by an additional noise component as the energy resolution of these pixels are comparable to those of the inner pixels. The excess is caused by the detection of K$_\upalpha$-fluorescence photons from cadmium \mbox{($E\approx23\,$k\eVd)} and tellurium \mbox{($E\approx27\,$k\eVd)} which originate from the CdTe crystal at the position of the guard ring. As expected, this excess disappears with an energy selection that excludes the CdTe fluorescence lines, see Fig.\,\ref{fig:am241} center.
   \item The line-like feature between (10/4) and (14/2) is caused by increased noise originating from a scratch on the crystal surface that happened during mounting operations. 
\end{itemize}

\subsection{$^\mathit{60}$Co and $^\mathit{137}$\hspace{-0.1em}Cs measurement}
\label{sec:high_erg}
In order to test the detector response for high energies, Caliste-O was set into its lowest gain mode and a $^{60}$Co and $^{137}$Cs source was used within the same acquisition. This double source measurement not only eases the energy calibration but also proofs the capability of a multi source detection, i.e. a source detection within a high background environment caused by Comptonization effects of other lines.

\begin{figure}[htb]
   \capstart
   \centering
   \includegraphics[width = 1.0\linewidth]{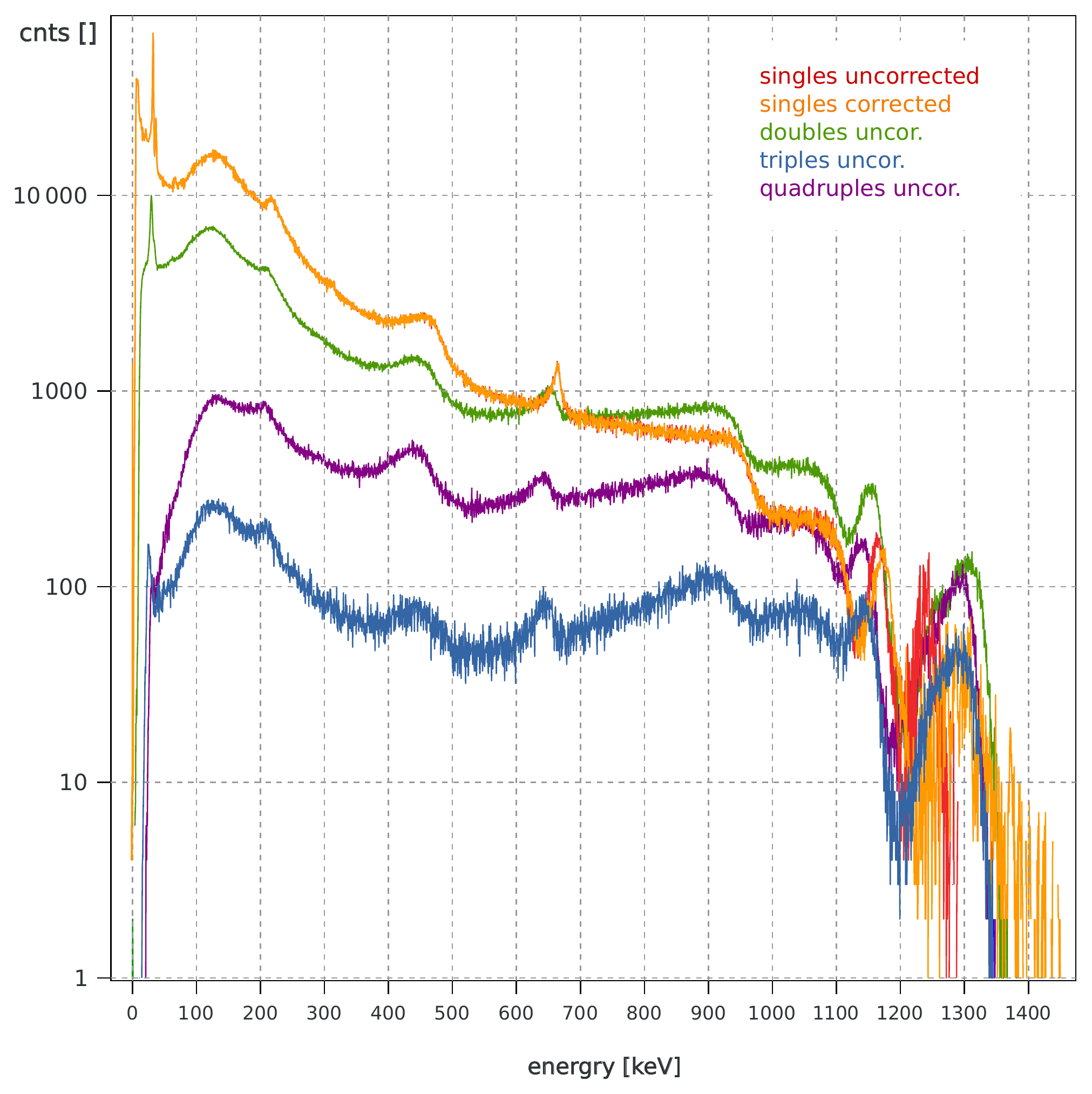}
   \caption{Spectra of a measurement with $^{60}$Co and $^{137}$Cs. The gain saturation was only corrected for the single events. Both $^{60}$Co peaks are clearly visible at 1.17\,MeV and 1.33\,MeV. See also Fig.\,\ref{fig:comptonization} for a model for the Comptonization.}
   \label{fig:Co60_spec}
\end{figure}

Figure \ref{fig:Co60_spec} shows the obtained spectra for all single pixel events, and for all split events into two, three, or four neighboring pixels (which are in the following named \textit{doubles}, \textit{triples}, and \textit{quadruples}, respectively). The single event spectrum shows a distortion effect near the end of its accessible energy range: the 1.17\,MeV $^{60}$Co peak is positioned 10\,keV too low while the 1.33\,MeV $^{60}$Co peak is shifted by $\sim\!\!100\,$keV to lower energies.

This effect is caused by a non-linear energy response near the saturation of the readout amplifier. While a hit map which is restricted to energies around the 1.17\,MeV peak shows a homogeneous counting rate, the hit map around the 1.33\,MeV peak is inhomogeneous as some of the readout channels are already saturated in their response. 

\noindent This saturation effect was corrected by multiplying the uncorrected (linearly calibrated) energies $E_\mathrm{old}$ with a function that approaches identity for low energies but which has a positive pole at the energy saturation $E_\mathrm{max}$:
\begin{equation}
	E_\mathrm{new} = E_\mathrm{old} \cdot \left(1 + \left(\frac{d}{E_\mathrm{max} - E_\mathrm{old}}\right)^6\right)
\end{equation}
The parameters for this correction (distortion $d$ and energy saturation $E_\mathrm{max}$) are obtained via a best fit to the two $^{60}$Co lines, while the model (pole of order 6) was chosen in order to minimize the fit error.
Because of the low counting statistics at high energies the parameters ($d$ and $E_\mathrm{max}$) have been determined globally for all pixels, which causes an additional energy uncertainty at the very end of the energy range, i.e. for $E > 1.2\,$MeV. Therefore, we define the energy range of \mbox{Caliste-O} to $2\,\mathrm{keV}\!<\!E\!<\!1.2\,$MeV, even though the 1.33\,MeV $^{60}$Co emission is clearly visible in the single event spectrum.

For the double, triple, and quadruple events a correction for energy saturation was not necessary because most of the split energies are at much lower energies than $E_\mathrm{max}$.
The energy resolution was measured with 
\begin{itemize}
	\item $\Delta E = 6.7$ keV FWHM at 662\,keV, (all single events)
	\item $\Delta E = 25.4$\,keV FWHM at 1.17\,MeV, (all single events)
	\item $\Delta E = 83.7$\,keV FWHM at 1.33\,MeV (all double events).
\end{itemize}

\subsection{Comptonization model}
\begin{figure}[tb]
   \capstart
   \centering
   \includegraphics[width = 1\linewidth]{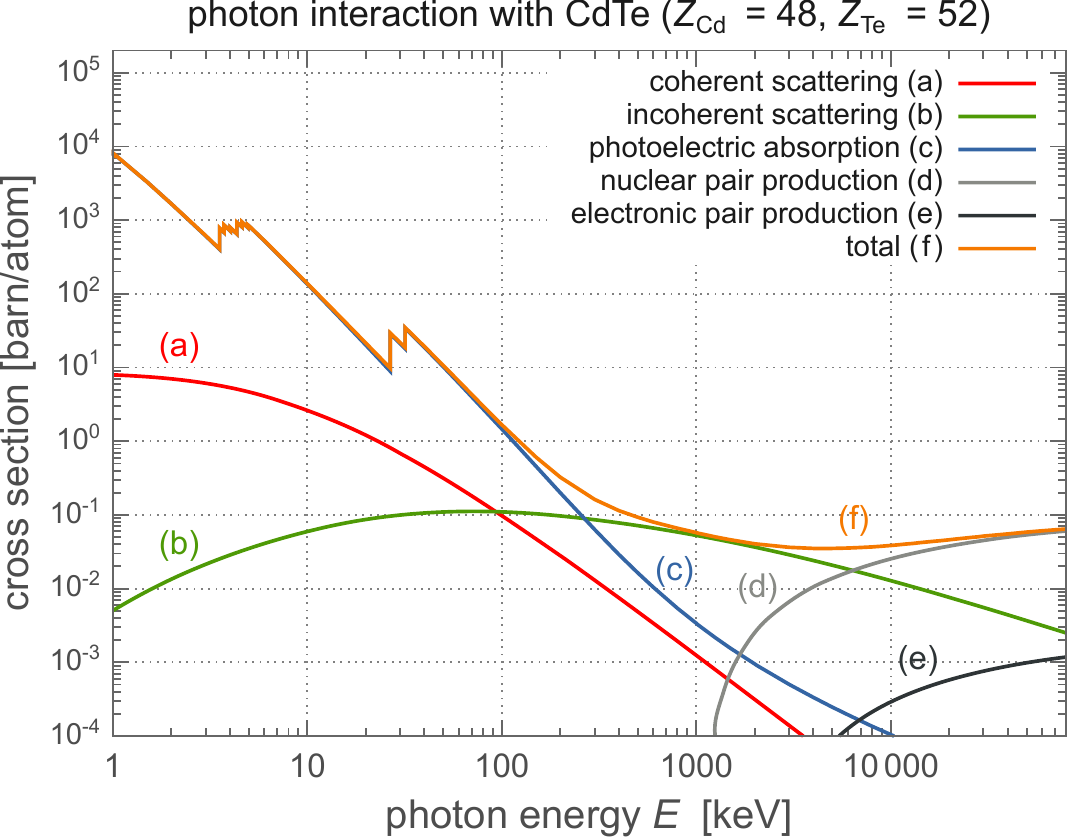} 
   \caption{Photon cross section for CdTe; data: XCOM Photon Cross Sections Database, see \cite{xcom-online}.}
   \label{fig:CdTe}
\end{figure}

Figure \ref{fig:CdTe} shows the cross section for the photon interaction with CdTe as a function of the photon energy. While for energies $E \lesssim 300$\,keV photoelectric absorption is the dominant interaction, incoherent scattering becomes dominant for higher energies. 
A proper understanding of the Compton interaction and its contribution to the background is therefore crucial for all measurements above several 100\,keV.

The Comptonization is composed of several structures that originate from the different interactions of the photons. The following scenarios were investigated in order to model the photon interactions:
\begin{itemize}
   \item The photon can be directly absorbed by the detector producing a peak at the photon energy.
   \item The photon can be incoherently scattered (Compton interaction) in the detector and thus deposits only a fraction of its energy according to the Klein-Nishina distribution \citep{Klein-Nishina}.
   \item The photon can be---one or several times---incoherently scattered in the materials around the detector. Afterwards, the scattered photon can interact with the detector by photoelectric absorption (full energy deposit of the scattered photon energy) or by another Compton interaction (fractional energy deposit).
\end{itemize}

\begin{figure*}[tbp]
   \capstart
   \centering
   \includegraphics[width = 1\linewidth]{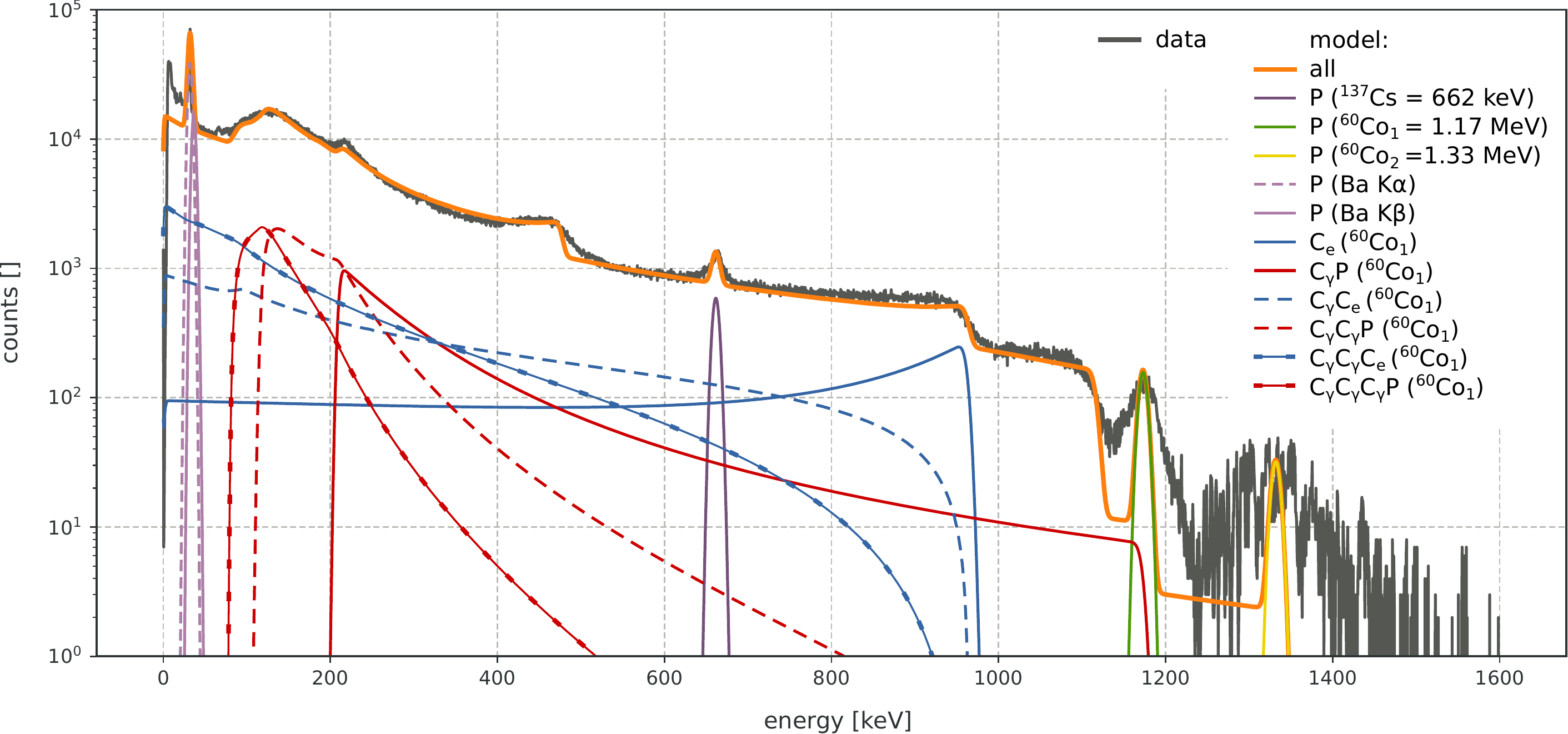}
   \caption{Single event spectrum of a $^{60}$Co and $^{137}$Cs measurement. The Comptonization is only shown for the $^{60}$Co emission at 1.17\,MeV. The label \textit{P} means \textit{photoelectric absorption}, $C_\gamma$ means \textit{Compton scattering outside the detector}, and $C_e$ means \textit{Compton scattering in the detector}. Example: $C_\gamma C_e$ means Compton scattering outside the detector followed by a Compton scattering in the detector; the scattered photon escapes the detector and the Compton electron is measured. The sum of the Comptonization models for the two $^{60}$Co emission lines and the $^{137}$Cs emission line fit the measured data very well.}
   \label{fig:comptonization}
\end{figure*}

The spectral contributions $S_i$ from these different scenarios have been computed for the emission lines of $^{137}$Cs~(662\,keV) and $^{60}$Co~(1.17 and 1.33\,MeV) and for the K$_\upalpha$~(32\,keV) and K$_\upbeta$~(36\,keV) emissions of barium (only photoelectric absorption). The resolution of the detector was included in the computation. The final spectrum is a linear combination of the different spectral contributions $S_i$: 
\begin{equation}
   S_\mathrm{data}=\sum_i a_i \, S_i
\end{equation}
where the $a_i$ coefficients are determined by fitting the linear combination to the spectrum with a least squares method.
Figure \ref{fig:comptonization} shows the result of this calculation compared to the spectrum. It shows also the different distributions corresponding to the different scenarios for the example of the 1.17\,MeV emission of $^{60}$Co. In the legend, P means \textit{energy deposited by photoelectric absorption by the detector}, C$_\mathrm{e}$ means \textit{energy deposited by Compton interaction in the detector}, and C$_\upgamma$ means \textit{Compton interaction anywhere but in the detector}.
The theoretical distribution fits to the spectrum everywhere except in three energy intervals:
\begin{itemize}
\item $E < 70$\,keV: fluorescence radiation of materials in the environment of the detector (i.e. copper) and multiple ($>\!3$) Compton interactions are not included in the model and cause an excess of the data compared to the model.
\item $E \approx 500$\,keV: the difference is probably caused by the optimization of the parameters $a_i$ to better fit the low energy response at $E < 70$\,keV.
\item $E > 1.1\,$MeV: the two peaks of $^{60}$Co show a reduced  energy resolution compared to the other line emissions. This effect is caused by the energy saturation of the readout channels, see discussion in Sect.\,\ref{sec:high_erg}.
\end{itemize}

\section{Discussion and outlook}
\label{sec:outlook}
The laboratory tests show that Caliste-O is fully functional with a homogeneous pixel response that matches the expectations. The accessible energy range was measured to be within $E_\mathrm{max,\,1} = 1.2\,$MeV for single events; this broad dynamic range allows to detect $^{60}$Co, which is of great interest as, for example, $^{60}$Co constitutes one of the dominant sources in the Fukushima Daiichi Plant \citep{Morichi}.

Source localization in the high energy range, i.e.~gamma-ray imaging, must be based on Compton imaging because of two reasons: first, the mask becomes transparent and, second, incoherent scattering becomes the dominant interaction within the crystal at high energies (see Fig.\,\ref{fig:CdTe}). Because Compton imaging results in a nearly $4 \uppi$ field-of-view heavy shielding of the camera is not foreseen, but can be easily added for specialized applications thanks to the compact design of the camera. For background rejection at low energies, a mask anti-mask imaging system\cite{anti-mask, anti-mask2} is foreseen in order to build a hybrid coded-mask/Compton camera.

Analyzing split events (doubles, triples, quadruples) shifts the detection limit to the double, triple, or quadruple value of $E_\mathrm{max,\,1}$, respectively. An $^{152}$Eu source (1.4\,MeV) detection is demonstrated with a 1\,mm CdTe CALISTE-HD in \cite{Dubos_2015}. An even enhanced detection efficiency in this energy range is expected with the 2\,mm CdTe CALISTE-O version. Its high energy response characterization using split and Compton events is ongoing and will be presented in a future paper.

A pixel individual characterization of the energy saturation effect is scheduled next in order to achieve an optimal response up to $E_\mathrm{max,\,1} = 1.2\,$MeV in the single event spectrum. This calibration will also enhance the quality of the multiple-event spectra. Furthermore, advanced analysis methods that include a quantitative measurement of the Comptonization processes will allow us to make precise flux estimations up to the MeV range and to enhance the multi-sources detection capabilities of Caliste-O. The proper understanding of all background features in combination with the spectroscopic capabilities will allow us to solve complex (multi-source) scenes in terms of source identification and localization. Following the presented initial detector characterization, the calculation and verification of accurate confidence interval estimations for source identification and source localization is scheduled next.

Integration and testing of Caliste-O into its new portable housing is scheduled for the end of this year.

\section*{Acknowledgement}

ORIGAMIX is a RSNR research program of \textit{Investissement d'Avenir}, referenced ANR-11-RSNR-0016, supported by the French Government and managed by the \textit{Agence Nationale de la Recherche}.

%The ORIGAMIX project is partially funded by the French Government, coordinated by the French National Research Agency as a part of the \textit{Investissements d'Avenir} Program, under reference ANR-11-RSNR-0016.

%% The Appendices part is started with the command \appendix;
%% appendix sections are then done as normal sections
%% \appendix

%% \section{}
%% \label{}

%% If you have bibdatabase file and want bibtex to generate the
%% bibitems, please use
%%

\section*{References}
\bibliographystyle{elsarticle-num}
\bibliography{lit}

%% else use the following coding to input the bibitems directly in the
%% TeX file.

%\begin{thebibliography}{00}

%% \bibitem[Author(year)]{label}
%% Text of bibliographic item

%\bibitem[ ()]{}

%\end{thebibliography}
\end{document}